\documentclass{emulateapj}
\usepackage{graphicx}
\def\r14{$R^{1/4}$}

\def\kms{\ifmmode{\mathrm{km\,s}^{-1}}\else{km\,s$^{-1}$}\fi}

\newcommand{\Vmax}{\ensuremath{V_\mathrm{max}}}
\newcommand{\VVmax}{\ensuremath{V/V_\mathrm{max}}}
\newcommand{\invVmax}{\ensuremath{1/V_\mathrm{max}}}
\newcommand{\meanVVmax}{\ensuremath{\langle V/V_\mathrm{max}\rangle}}
\newcommand{\thetaL}{\ensuremath{\theta_\mathrm{L}}}

\def\HST{{\it HST}}

\newcommand{\meanmue}{\ensuremath{\langle\mu\rangle_\mathrm{e}}}
\def\Pnull{$P_{\rm null}$}
\def\reff{\ifmmode{R_{\rm e}}\else{$R_{\rm e}$}\fi}
\def\mueff{\ifmmode{\mu_{\rm e}}\else{$\mu_{\rm e}$}\fi}
\def\mueffH{\ifmmode{\mu_{{\rm e},H}}\else{$\mu_{{\rm e},H}$}\fi}
\def\mueffK{\ifmmode{\mu_{{\rm e},K}}\else{$\mu_{{\rm e},K}$}\fi}
\def\LKBul{\ifmmode{L_{K,\mathrm{Bul}}}\else{$L_{K,\mathrm{Bul}}$}\fi}
\def\LKDis{\ifmmode{L_{K,\mathrm{Disk}}}\else{$L_{K,\mathrm{Disk}}$}\fi}
\def\LKSun{\ifmmode{L_{K,\odot}}\else{$L_{K,\odot}$}\fi}
\def\Ieff{\ifmmode{I_{\mathrm{e}}}\else{$I_{\mathrm{e}}$}\fi}
\def\IeffK{\ifmmode{I_{\mathrm{e},K}}\else{$I_{\mathrm{e},K}$}\fi}
\def\IZeroK{\ifmmode{I_{\mathrm{0},K}}\else{$I_{\mathrm{0},K}$}\fi}
\def\IZeroSer{\ifmmode{I_{0,\mathrm{Ser}}}\else{$I_{0,\mathrm{Ser}}$}\fi}
\def\IZeroSerK{\ifmmode{I_{0,\mathrm{Ser},K}}\else{$I_{0,\mathrm{Ser},K}$}\fi}
\def\IZeroBulK{\ifmmode{I_{0,\mathrm{Bul},K}}\else{$I_{0,\mathrm{Bul},K}$}\fi}
\def\IZeroBul{\ifmmode{I_{0,\mathrm{Bul}}}\else{$I_{0,\mathrm{Bul}}$}\fi}
\def\muZeroSer{\ifmmode{\mu_{0,{\rm Ser}}}\else{$\mu_{0,{\rm Ser}}$}\fi}
\def\muZeroSerK{\ifmmode{\mu_{0,{\rm Ser},K}}\else{$\mu_{0,{\rm Ser},K}$}\fi}
\def\muZeroBul{\ifmmode{\mu_{0,{\rm Bul}}}\else{$\mu_{0,{\rm Bul}}$}\fi}

\def\IZeroDis{\ifmmode{I_{0,\mathrm{Disk}}}\else{$I_{0,\mathrm{Disk}}$}\fi}
\def\IZeroDisK{\ifmmode{I_{0,\mathrm{Disk},K}}\else{$I_{0,\mathrm{Disk},K}$}\fi}
\def\muZeroDis{\ifmmode{\mu_{0,{\rm Disk}}}\else{$\mu_{0,{\rm Disk}}$}\fi}
\def\muZeroDisK{\ifmmode{\mu_{0,{\rm Disk},K}}\else{$\mu_{0,{\rm Disk},K}$}\fi}
\def\muZeroDis{\ifmmode{\mu_{0,{\rm Disk}}}\else{$\mu_{0,{\rm Disk}}$}\fi}

\newcommand{\MagKBul}{\ensuremath{M_{K,\mathrm{Bul}}}}
\newcommand{\MagKDis}{\ensuremath{M_{K,\mathrm{Disk}}}}

\newcommand{\LumKBul}{\ensuremath{L_{K,\mathrm{Bul}}}}
\newcommand{\LumKDis}{\ensuremath{L_{K,\mathrm{Dis}}}}

\newcommand{\LumPt}{\ensuremath{L_{\mathrm{PS}}}}

\shorttitle{Scaling relations of bulges}
\shortauthors{Balcells, Graham, \& Peletier}

\begin{document}

\doublespace 

\title{Galactic bulges from {\sl Hubble Space Telescope} NICMOS observations:  global scaling relations\altaffilmark{1}$^{,}$\altaffilmark{2}}
\author{Marc Balcells}
\affil{Instituto de Astrof\'\i sica de Canarias, 38200 La Laguna, Tenerife, Spain}
\author{Alister W. Graham\altaffilmark{3}}
\affil{Centre for Astrophysics and Supercomputing, Swinburne University of Technology, Hawthorn, Victoria 3122, Australia}
\and
\author{Reynier F. Peletier\altaffilmark{4}}
\affil{Kapteyn Institute, University of Groningen, Postbus 800, 9700 AV Groningen, The Netherlands}
\email{balcells@iac.es}

\slugcomment{ApJ accepted 2007 April 27} 

\altaffiltext{1}{Based on observations made with the NASA/ESA {\sl 
Hubble Space Telescope}, obtained at the Space Telescope Science 
Institute, which is operated by the Association of Universities for 
Research in Astronomy, Inc., under NASA contract NAS 5-26555.}
\altaffiltext{2}{Based on observations made with the Isaac Newton and William Herschel Telescopes operated on the island of La Palma by the Isaac Newton Group of Telescopes in the Spanish Observatorio del Roque de los Muchachos of the Instituto de Astrof\'\i sica de Canarias} 
\altaffiltext{3}{Also: Instituto de Astrof\'\i sica de Canarias, 38200 La Laguna, Tenerife, Spain}
\altaffiltext{4}{Also: School of Physics and Astronomy, University of Nottingham, NG7 2RD, UK}

\begin{abstract}
We investigate bulge and disk scaling relations using a volume-corrected
sample of early- to intermediate-type disk galaxies in which, importantly,
the biasing flux from additional nuclear components has been modeled and removed. Structural parameters are
obtained from a seeing-convolved, bulge$+$disk$+$nuclear-component
decomposition applied to near-infrared surface brightness
profiles spanning $\sim$10 pc to the outer disk.  
Bulge and disk parameters, and bulge-to-disk ratios, are analyzed as a
function of bulge luminosity, disk luminosity, galaxy central velocity dispersion, and galaxy Hubble type.  
Mathematical expressions are given for the stronger relations, which can be used to
test and constrain galaxy formation models.
Photometric parameters of both bulges and disks are observed to correlate with bulge luminosity and with central velocity dispersion.  
In contrast, for the unbarred, early to intermediate types covered by the sample, Hubble type does not correlate with bulge and disk components, nor their various ratios.  
In this sense, the early-to-intermediate spiral Hubble sequence is scale-free.  
However, galaxies themselves are not scale-free, the critical scale being the luminosity of the bulge.  
Bulge luminosity is shown to affect the disk parameters, such that central surface brightness becomes fainter, and scale-length bigger, with bulge luminosity.  
The lack of significant correlations between bulge pararmeters (size, luminosity or density)  on disk luminosity, remains a challenge for secular evolution models of bulge growth.  

\end{abstract}


\section{Introduction}
\label{Sec:Introduction}

In a companion paper (Balcells, Graham, \& Peletier 2007, hereafter Paper~III), we present high-resolution near-infrared (NIR) surface brightness profiles of early- to intermediate-type disk galaxies.  These are derived from {\sl Hubble Space Telescope} (\HST) NICMOS images and extended with NIR ground-based profiles from UKIRT images.  That paper provides profile decompositions into bulge, disk, and nuclear components, and analyzes the nuclear properties of the bulges.  In the present paper, we use the profile decompositions of Paper~III to analyze the global scaling relations for bulges and disks, free from the biasing influence of additional nuclear components.

Global scaling relations provide useful diagnostics on the structure of disk galaxies.  For bulge components, comparison with the scaling relations of elliptical galaxies shows to what degree bulges are either similar or different to ellipticals.  Whether bulges resemble spheroids or disks is still an open question (Wyse, Gilmore, \& Franx 1997; Kormendy \& Kennicutt 2004).  To give one example of the complexity of this issue, some apparent bulges in face-on, barred S0 galaxies are in fact inner disks (Kormendy 1993; Erwin et al.\ 2003), which of course has implications for galaxy formation models.  For both bulges and the outer, large-scale disks, the scaling relations may provide clues on the long-standing question of the origin of the Hubble sequence. 

The sample of de Jong \& van der Kruit (1994) has provided a useful reference for intermediate- to late-type field disk galaxies, and its scaling relations have been analyzed by de Jong (1996) and Courteau, de Jong, \& Broeils (1996).  Graham (2001, hereafter G01) and  MacArthur, Courteau, \& Holtzman (2003) provide a more recent study of intermediate- to late-type spirals, using S\'ersic fits for the bulge.  Hunt, Pierini, \& Giovanardi (2003, hereafter HPG03) have also analyzed bulge and disk scaling relations from a sample of disk galaxies in the Perseus-Pisces supercluster;  their sample comprises mostly Sb and later-type spirals.  At the other end of the Hubble sequence, S0 galaxies are often discussed together with  ellipticals, but are rarely compared to spiral galaxies.   The study presented in this paper is complementary to the ones above as it samples early- to intermediate-type, S0 through Sbc, disk galaxies.  Moreover, the analysis presented here uses \HST\ images to subtract the flux of additional nuclear components such as star clusters and nuclear disks, which have biased previous studies of bulge, and hence disk, parameters.  This is perhaps best evidenced through the reporting of {\r14}-like bulges with S\'ersic indices $n\sim 4-6$ when using low-resolution ground-based data which smear out the flux from the unresolved nuclear components (e.g., Andredakis, Peletier, \& Balcells 1995, hereafter APB95).  Higher-resolution studies with \HST\ have since revealed that the majority of such galaxies have noticeably less concentrated bulges, with $n \lesssim 3$, but clear additional components (Balcells et al.~2003, hereafter Paper~II; Paper~III).  

Bulge and disk scaling relations have traditionally been studied as a function of Hubble type, commonly parametrized with the revised type index T (de Vaucouleurs et al.\ 1991, hereafter RC3).  The lack of a correlation between the ratio of the bulge effective radius $\reff$  and the disk scale length $h$ with Hubble type, led to the statement that the spiral Hubble sequence is scale-free (de Jong 1996; Courteau et al.\ 1996).  In a scale-free situation, an 'ice-berg' model, in which bulge surface brightness rather than effective radius determines how much the bulge protrudes above the disk, might explain the higher prominence of earlier-type bulges (G01).  
That the Hubble sequence is scale-free has implications for bulge formation models, as secular evolution models (Pfenniger \& Norman 1990; Norman, Sellwood, \& Hasan 1996) predict that bulge and disk scale lengths are correlated (Combes et al.\ 1990).  

The parameters used for the scaling relations, derived from NIR images, are less affected by dust extinction than those derived from optical images.  Furthermore, $M/L$ variations with population age and metallicity are small in the NIR, yielding small differences between the photometric length-scales and the stellar-mass scales of the galaxies.  
We will show that bulge luminosity is indeed a key yardstick that traces the values of bulge, disk, and global galaxy parameters.  Section~\ref{Sec:Data} describes the sample and the profile decomposition.  Section~\ref{Sec:Completeness} discusses selection biases and provides a volume correction.  The main results of the paper are contained in Figures \ref{Fig:BDparMK}, \ref{Fig:BDparDisMK}, \ref{Fig:BDparSIG}, and \ref{Fig:BDparT}, and analyzed in Section~\ref{Sec:GlobalParameters}, for bulges (\S\ref{Sec:BulgeParameters}), disks (\S\ref{Sec:DiskParameters}), and bulge-to-disk scaling relations (\S\ref{Sec:BDscaling}).  We discuss two specific issues in Section~\ref{Sec:Discussion}: the trends of the galaxy central surface brightness with spheroid luminosity (\S\,\ref{Sec:DiscussionMu0}), and the scale-free nature of the Hubble sequence (\S\,\ref{Sec:DiscussionHubbleSequence}). 
A Hubble constant of $H_{0} = 75$ km\,s$^{-1}$\,Mpc$^{-1}$ is used throughout.  

\section{Galaxy sample and data}
\label{Sec:Data}

The galaxy sample is described in detail in Paper~III.  It comprises 19 galaxies of types S0-Sbc extracted from the diameter-limited sample of inclined galaxies of Balcells \& Peletier (1994).  
The diameter limit puts the selected galaxy diameters in the broad range 15 kpc to 80 kpc.  The selection excluded very dusty bulges, so the sample has a slight bias toward quiescent bulges.  
Basic properties of the sample are listed in Table\,1 of Paper~III.  The 19 galaxies were observed with NICMOS on \HST\  through the F160W filter.  Data reduction is described in Peletier et al.\ (1999, hereafter Paper~I).  The derivation of surface brightness profiles is described in Paper~III.  $K$-band surface brightness profiles obtained from UKIRT images (APB95; Peletier \& Balcells 1997) were scaled to the $H$-band and linked smoothly to the \HST\  profiles, yielding surface brightness profiles which span from $\sim$20 pc to several kpc, thus covering the nucleus, bulge, and disk-dominated region of each galaxy.  

We performed a one-dimensional profile decomposition using an exponential model for the disk and S\'ersic (1963) model (Graham \& Driver 2005) for the bulge, 

\begin{equation}
I(R) = I(0)\,\exp\{-b_n\,(R/\reff)^{1/n}\},
\label{Eqn:Sersic}
\end{equation}

\noindent where $\reff$ encloses half the model light, $n$ measures the curvature of the profile, and $b_n\approx 1.9992n-0.3271$.    The decomposition is described in detail in Papers~II and III.  Due to the presence of positive nuclear residuals in some galaxies, unresolved (PS) or resolved (exponential) components were added to the fitting function.  The properties of these nuclear components are analyzed in Paper~III, which also shows the profiles, the fits, and lists the best-fit parameters (Tables\,2--4 of that paper).  Those parameters provide the basis for the analysis of the scaling relations presented here.  Our analysis ignores galaxian sub-components that may be found outside the bulges, such as rings and lenses (Prieto et al.\ 2001).  We also ignore bar components.  Our sample selection took galaxies listed as unbarred, although the presence of bars in some of the more edge-on cases cannot be ruled out.  

\subsection{Sample completeness}
\label{Sec:Completeness}

Given the small sample size ($N=19$) and the various selection processes involved, we have checked to what degree our sample is a fair representation of the local S0-Sbc galaxy population.  We follow the standard \VVmax\ formalism developed by Thuan \& Seitzer (1977).  Briefly, for a diameter-limited sample, under the assumption that surface brightness is independent of distance, a galaxy's angular diameter $\theta$ is inversely proportional to its distance, and it is straightforward to compute the maximum distance $d_\mathrm{max}$ at which the galaxy would still be included in a sample limited by $\theta \geq \thetaL$, with $\thetaL = 2\arcmin$ in our case.  
We then compute the volume $V$ of a sphere of radius equal to the distance $d$ to the galaxy, and the volume $V_\mathrm{max}$ of the sphere out to $d_\mathrm{max}$.  The ratio of volumes is \VVmax\ $= (\thetaL/\theta)^3$.  For objects randomly distributed in space, \VVmax\ should be uniformly distributed between 0 and 1, with a mean \meanVVmax\ $= 0.5 \pm 1/\sqrt(12\,N)$, where $N$ is the number of objects.  

For the original sample selected in Balcells \& Peletier (1994), comprising 43 objects, the UGC red diameters yield \meanVVmax\ $=0.480 \pm 0.044$, hence that sample is statistically complete.  The sample of 19 objects imaged with \HST\  has \meanVVmax\ $=0.413 \pm 0.066$.  
This shows that the \HST\ sample is mildly biased toward nearby objects; this reflects the selection of targets for \HST\  imaging, which, other things being equal, favored high spatial resolution.  The bias introduced is small, in any case.  

By weighting each galaxy by \invVmax, our diameter-limited sample mimics a volume-limited sample.
The  weighting is essential whenever mean values or volume-related quantities are sought; we have also explored its use when computing scaling relations between various galaxy parameters, see \S\ref{Sec:GlobalParameters}.  \Vmax\ is given by 

\begin{equation}
V_\mathrm{max} = \frac{4\pi}{3}(d_\mathrm{max})^3 = \frac{4\pi}{3}\left(\frac{d\times\theta}{\thetaL}\right)^3.
\end{equation}

The distribution of \invVmax\ with bulge $K$-band absolute magnitude will be shown in Figure\,\ref{Fig:BDparMK}j.

\begin{figure*}
\begin{center}
\caption{\label{Fig:BDparMK}
The dependence of bulge and disk parameters, and bulge-disk ratios, on the $K$-band bulge absolute magnitude.  All parameters are from Table~3 in Paper~III, except for the velocity dispersions which are from Falc\'on-Barroso et al.\ (2002), and are listed in Table~1 in Paper~III.  \textit{Solid lines} show orthogonal regressions to the data points; the corresponding relations are given in \S\ref{Sec:GlobalParameters}.  
The \textit{open circle} corresponds to NGC~5577, an outliner in most of the distributions;  it has been excluded when computing all of the regressions.  
   ({\it a}) Effective radius of the (S\'ersic) bulge component. The dashed line shows $L \sim \reff^2$, offset for clarity.  
   ({\it b}) Effective surface brightness of the S\'ersic component.  
   ({\it c}) Extrapolated $H$-band central surface brightness of the bulge component.  
   ({\it d}) Disk major-axis scale length. 
   ({\it e}) Face-on $H$-band extrapolated disk central surface brightness.  The horizontal dotted line is the canonical Freeman value, using the mean color for the sample $B-H=3.7$.  
   ({\it f}) Central velocity dispersion.  The dashed line gives the Faber-Jackson relation for Coma ellipticals, from Pahre et al.\ (1998).
   ({\it g}) Bulge-to-disk central brightness ratio $\log(I_{0,B}/I_{0,D})$.
   ({\it h}) Ratio $\reff/h$ between the bulge effective radius and the disk scale length.
   ({\it i}) Bulge-to-disk luminosity ratio $B/D$.
   ({\it j}) \invVmax\ volume correction, normalized to the maximum \invVmax.}   
\end{center}
\end{figure*}

\addtocounter{figure}{-1}

\begin{figure*}[htbp]
\begin{center}
\includegraphics[height=16cm]{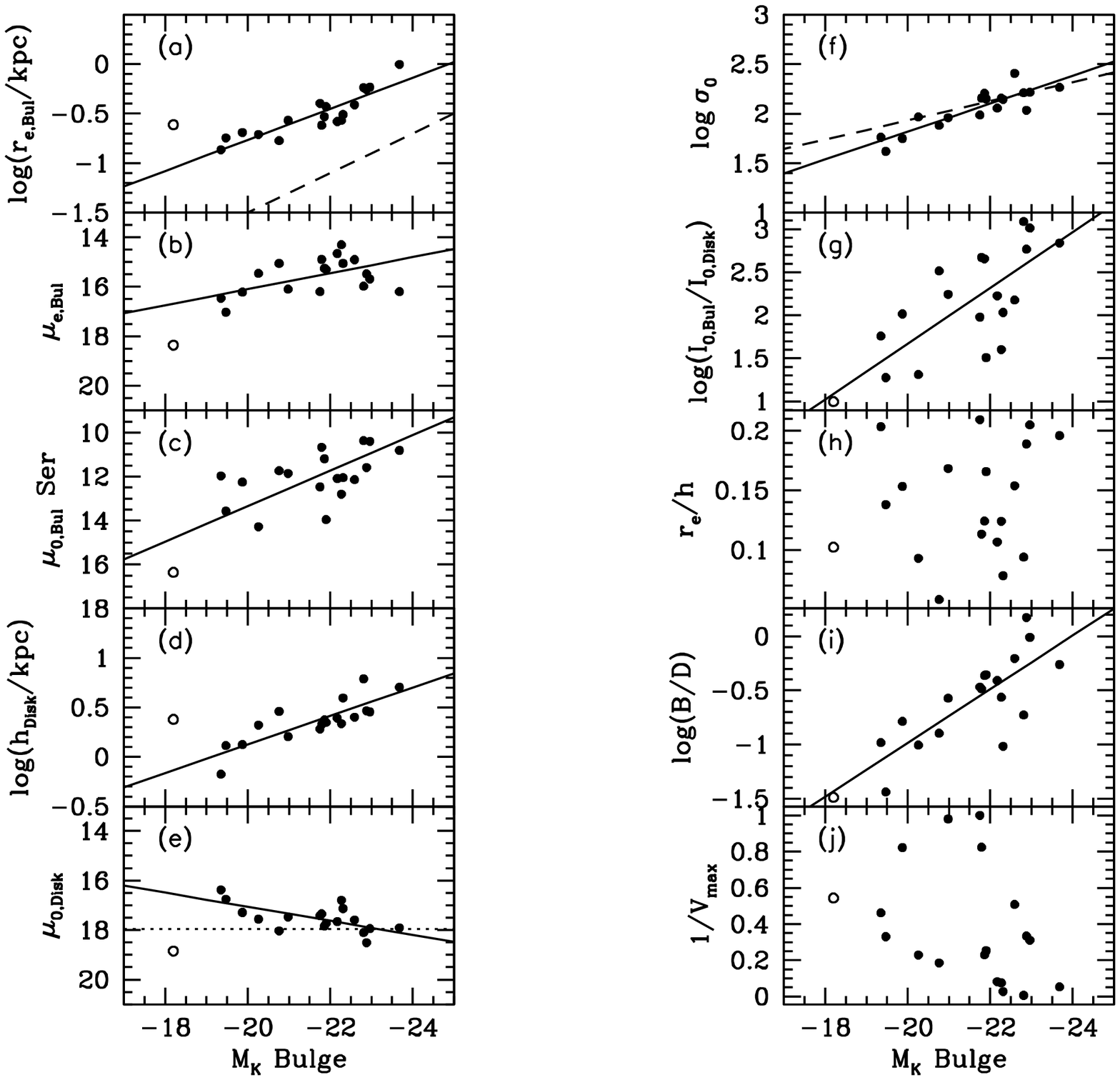}
\caption{See caption in previous page.}
\label{default}
\end{center}
\end{figure*}

\begin{figure*}[htbp]
\begin{center}
\includegraphics[height=16cm]{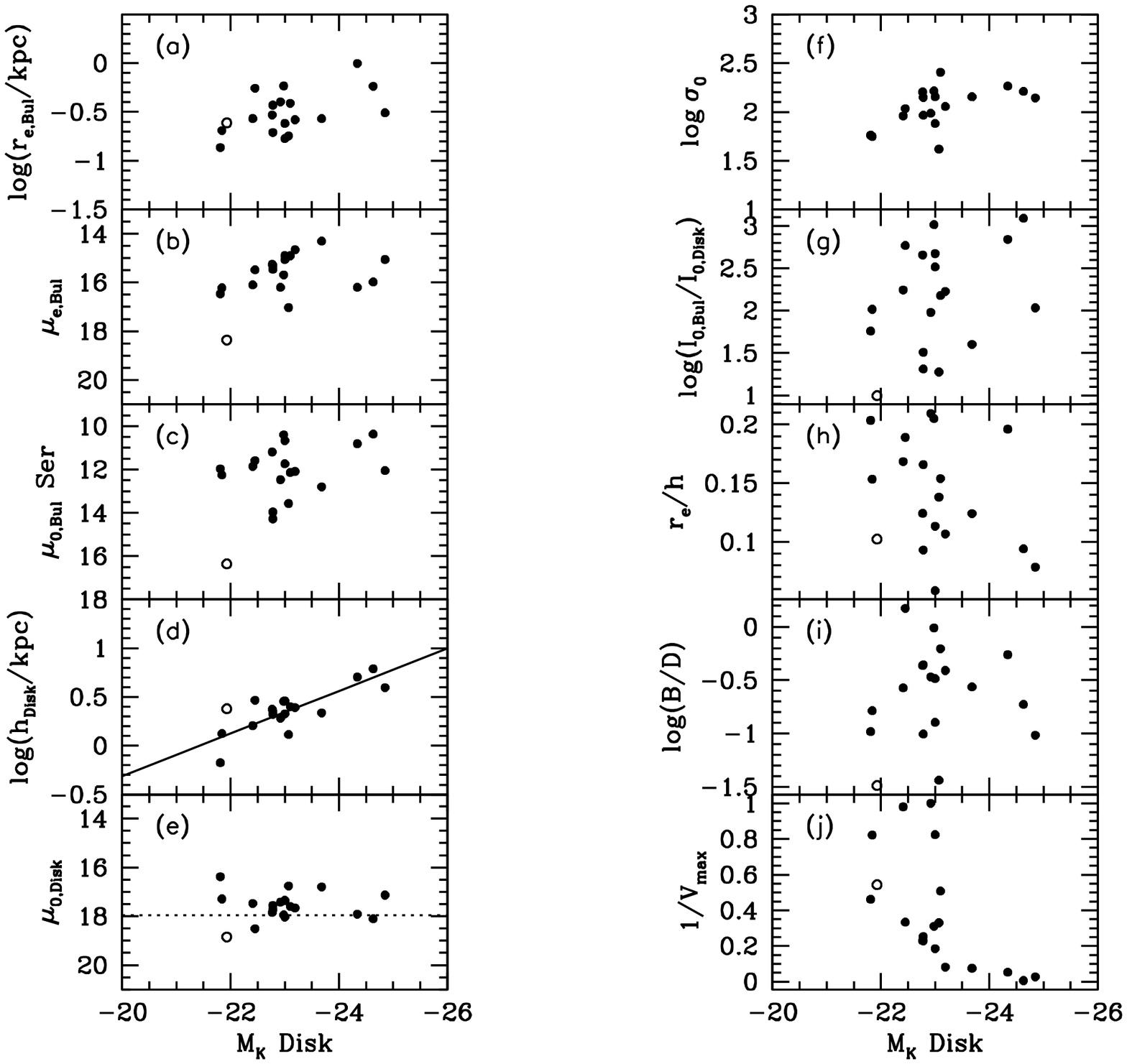}
\caption{\label{Fig:BDparDisMK}
Identical to Figure\,\ref{Fig:BDparMK}, except for the use of the $K$-band disk absolute magnitude in the abscissae.}
\end{center}
\end{figure*}

\begin{figure*}[htbp]
\begin{center}
\includegraphics[height=16cm]{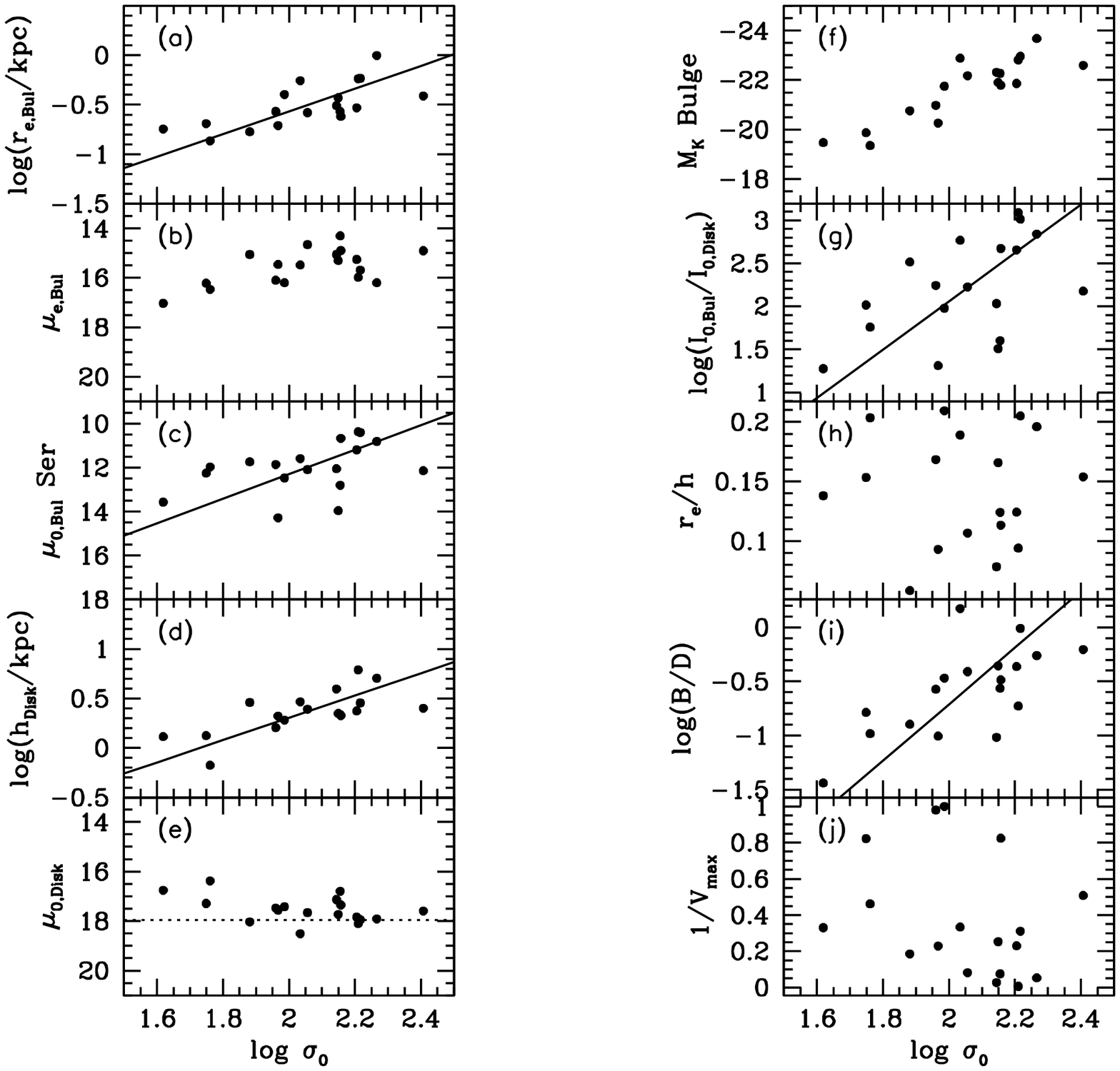}
\caption{\label{Fig:BDparSIG}
Identical to Figure\,\ref{Fig:BDparMK}, except for the use of the central velocity dispersion in the abscissae.}
\end{center}
\end{figure*}

\begin{figure*}[htbp]
\begin{center}
\includegraphics[height=16cm]{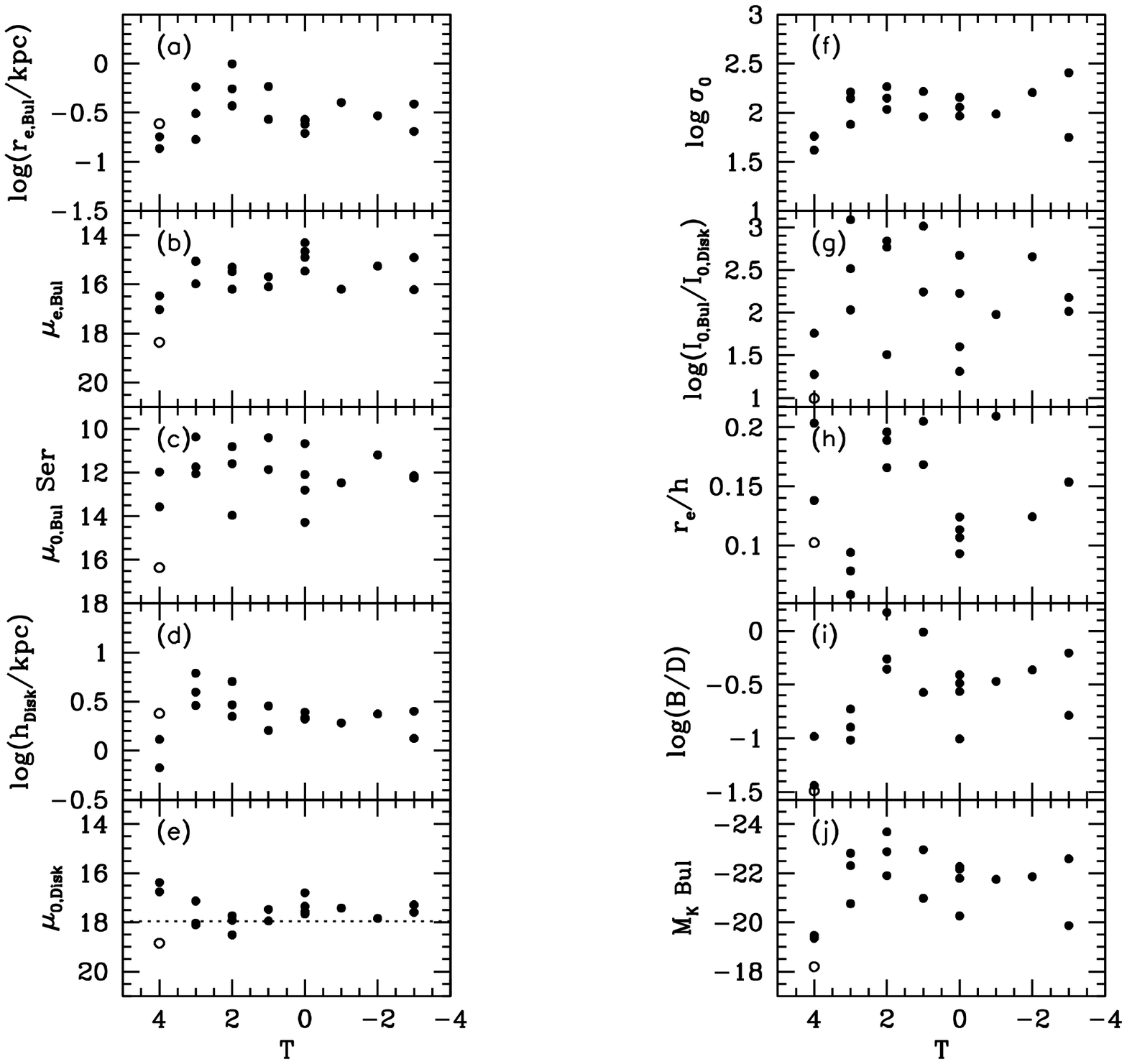}
\caption{\label{Fig:BDparT}
Identical to Figure\,\ref{Fig:BDparMK}, except that T is used in the abscissae, and 
panel (j) shows T vs.~bulge absolute magnitude.  }
\end{center}
\end{figure*}

\begin{figure}[htbp]
\begin{center}
\includegraphics[width=0.65\textwidth, angle=-90]{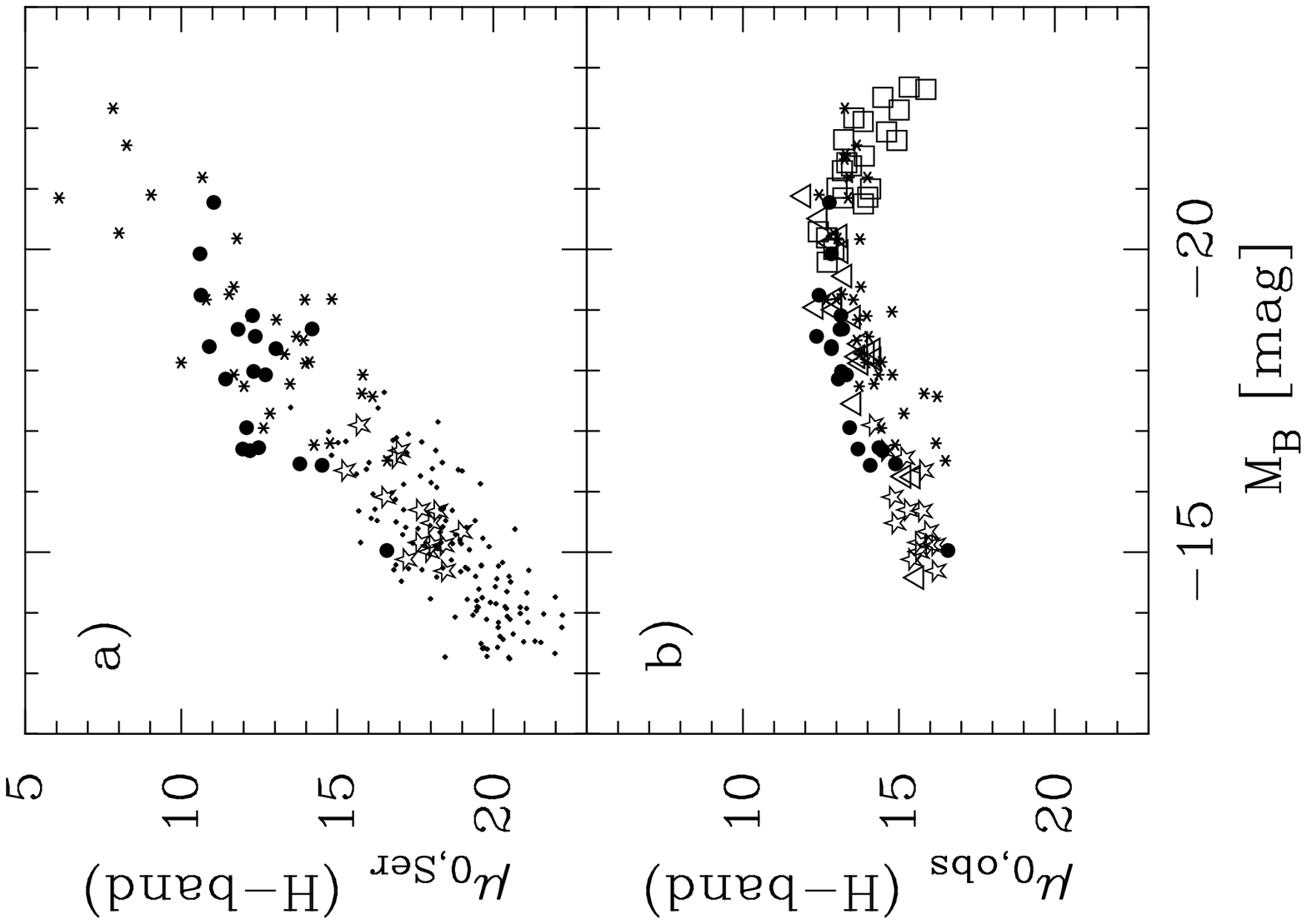}
\caption{\label{Fig:mu0MB} The central $H$-band surface brightness of spheroids vs. their $B$-band absolute magnitude.  ({\it a}) Extrapolated central surface brightness of the S\'ersic fit to the profile.  ({\it b}) Measured surface brightness at 50 pc from the center.  All values are corrected for Galactic extinction, cosmological dimming and K-correction.  
    {\it Filled circles}: bulges, this work.  
    {\it Asterisks}:  ellipticals from Caon et al.\ (1993).
    {\it Triangles}:  "power-law" ellipticals from L95.
    {\it Open squares}: "core" ellipticals from L95.  
    {\it Open stars}: Coma dwarf ellipticals from Graham \& Guzm\'an (2003).  
    {\it Points}:  Virgo dwarf ellipticals from Binggeli \& Jerjen (1998).  See footnote 7 for conversions.}
\end{center}
\end{figure}

\begin{figure*}[htbp]
\begin{center}
\includegraphics[width=0.95\textwidth]{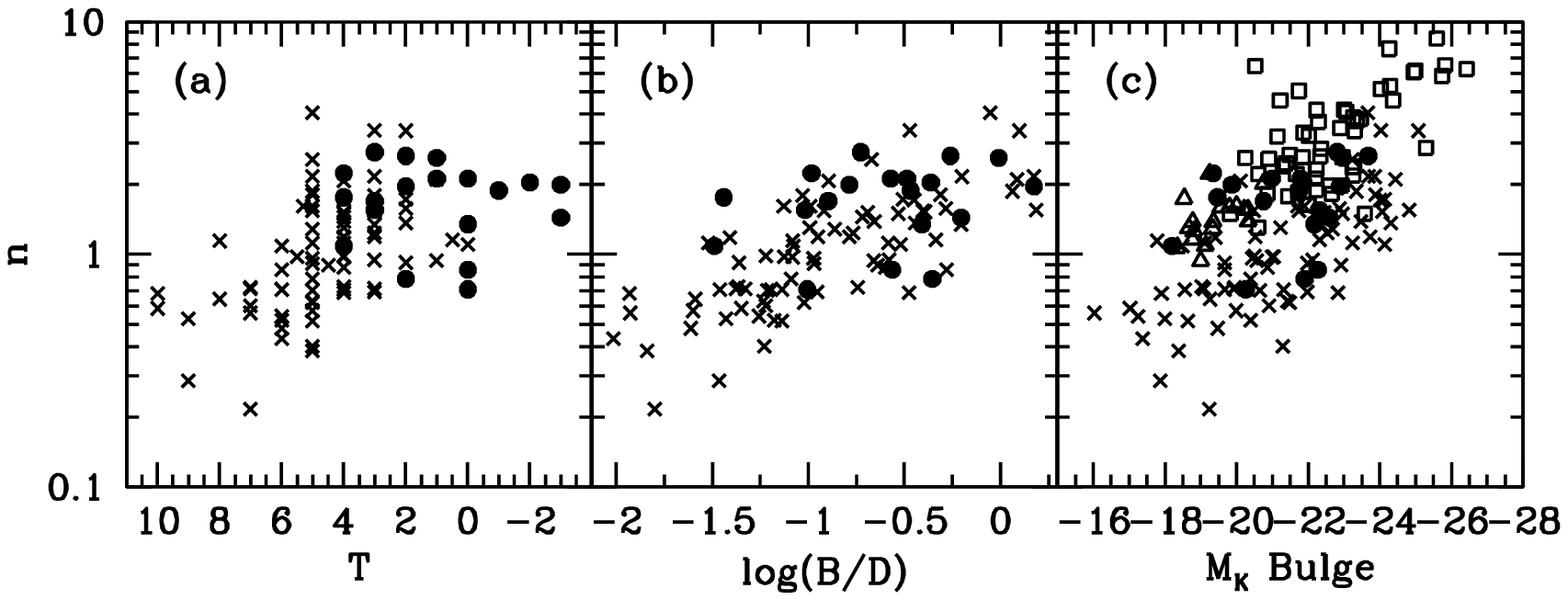}
\caption{The bulge S\'ersic index $n$ is plotted against: 
    ({\it a}) the revised morphological type index T from the RC3;
    ({\it b}) $B/D$derived from the best-fit parameters; and
    ({\it c}) the bulge $K$-band absolute magnitude derived from $B/D$and the galaxy $K$-band absolute magnitude, corrected for Galactic extinction, cosmological dimming and K-correction.
    {\it Filled circles}: bulges, this work.  
    {\it Crosses}:  bulges from the de Jong \& van der Kruit (1994) sample, as analyzed by G01.
    {\it Triangles}: Coma dwarf ellipticals from Graham \& Guzm\'an (2003).
    {\it Squares}: Virgo ellipticals from Caon et al.\ (1993).}
\label{Fig:NvsTvsBDvsMK}   
\end{center}
\end{figure*}

\section{Global parameters}
\label{Sec:GlobalParameters}

This section presents global scaling relations inferred from the bulge-disk decompositions.  Galaxy parameters are corrected for foreground Galactic extinction (Schlegel, Finkbeiner, \& Davis 1998), $(1+z)^4$ cosmological dimming, and $K$-correction (Poggianti 1997).   The disk parameters have been corrected to a face-on aspect, assuming transparent disks.  
The corrected parameters are listed in Table~3 of Paper~III.  
Figures\,\ref{Fig:BDparMK}--\ref{Fig:BDparT} display the dependencies of global parameters on bulge absolute magnitude, disk absolute magnitude, central velocity dispersion, and galaxy Hubble type, respectively.  Strong correlations with bulge absolute magnitude are generally found, with one single deviant point at the faint end.  This corresponds to NGC~5577 (Sbc), and we exclude this galaxy from the regression analysis in order not to bias the global trends.  
We computed the scaling relations via unweighted orthogonal regressions, following the algorithm of York (1966).  
We explored weighted orthogonal regressions using \invVmax\ weights: resulting fitting coefficients changed by typically $\sim$10\%, but a few of the weighted regressions failed to converge, hence, 
for the sake of a uniform treatment, the parameters presented in the following subsections are those of the unweighted regressions.   For each correlation we list the Spearman rank-order correlation coefficient ($S_\mathrm{R}$) and the probability of a null correlation (\Pnull).  

\subsection{Bulge parameters}
\label{Sec:BulgeParameters}

\subsubsection{Sizes}
\label{Sec:Sizes}

Bulge effective radii (Fig.\,\ref{Fig:BDparMK}a) range from 0.1 to 1 kpc.  They increase with bulge luminosity, as found for ellipticals (e.g.\ Hubble 1926; Binggeli, Sandage, \& Tarenghi 1984, hereafter VCC).  An orthogonal regression to the $\reff$--$\LKBul$ relation gives\footnote{$M_{K,\odot} = 3.41$ (Allen 1973)}

\begin{equation}
\reff/{\rm kpc} = 10^{-0.52 \pm 0.04}( \LKBul/10^{10}\LKSun)^{0.39 \pm 0.06}
\label{Eqn:reMK}
\end{equation}

\noindent ($S_\mathrm{R} = -0.87$; \Pnull\ $= 3.0\cdot 10^{-4}$).  
A strong correlation exists also between $\reff$ and bulge central velocity dispersion (Fig.\,\ref{Fig:BDparSIG}a; $S_\mathrm{R} = 0.74$; \Pnull\ $= 2.3\cdot 10^{-3}$).  
In contrast, $\reff$ shows no dependence on T (Fig.\,\ref{Fig:BDparT}a; $S_\mathrm{R} = -0.17$; \Pnull\ $= 0.46$).  Both of these results are similar to those of HPG03.  The correlation of $\reff$ with \textsl{disk} absolute magnitude is also null (Fig.\,\ref{Fig:BDparDisMK}a; $S_\mathrm{R} = -0.32$; \Pnull\ $= 0.18$).

M\"ollenhoff \& Heidt (2001) report $\reff \sim L^{0.84}$ for their bulge sample.  Part of the difference between the exponents appears to be due to the type of regression performed, as an \textsl{orthogonal} regression to their $K$-band data yields $\reff\sim L^{0.51 \pm 0.06}$, i.e., at 2-$\sigma$ from our result.  Our two samples cover similar bulge luminosity ranges, but their $\reff$ go up to 10 kpc for the largest bulges.  Their fits, using ground-based data, include any nuclear components as part of the bulge.  Given the 90\% detection frequency of nuclear components in our sample, such components are likely to be present in their galaxies, which may boost S\'ersic indices up to $n\sim 6$ (Paper\,II); and such biases yield higher values of $\reff$.  

For elliptical galaxies, VCC show that the  $\reff \sim L^\alpha$ relation has $\alpha < 0.5$ for low-luminosity ellipticals ($M_B + 5\,\log h_{50} \ga -20.5$), and $\alpha > 0.5$ for giant ellipticals ($M_B + 5\,\log h_{50} \la -20.5$).  Our bulges have absolute blue magnitudes in the range $-14<M_B + 5\,\log\,h_{50} <-20$ \footnote{Given a mean color of $B-K=4.0$ for our bulges (Peletier \& Balcells 1997).}.  Over that range, the $\reff - L$ relation for VCC ellipticals is approximately $\reff \sim L^{0.45}$.  Comparing with eqn.\,\ref{Eqn:reMK}, we conclude that the size--luminosity relations for bulges has similar slope to that of ellipticals of the same luminosity.  

\subsubsection{Effective surface brightness}
\label{Sec:MuEff}

The mean effective surface brightness is readily obtained from $\reff$ as \meanmue\ = $K +5\,\log\,\reff + 2.5\,\log\,2\pi$, where $K$ is the total apparent $K$-band magnitude of the bulge.  VCC showed that, for faint ellipticals, \meanmue\ becomes brighter with luminosity, while it becomes fainter with increasing luminosity for giants.    For bulges, the $\reff - L$ relation (eqn.\,\ref{Eqn:reMK}) indicates that \meanmue\ brightens with bulge luminosity over the bulges' luminosity range.  The $\reff\sim L^{0.5}$ which corresponds to \meanmue\ being independent of luminosity isshown with an offset dashed line in Fig.\,\ref{Fig:BDparMK}a). 

The effective surface brightness, $\mueff \equiv \mu(\reff) = -2.5\,\log\,\Ieff $, is shown against $K$-band absolute bulge magnitude in Figure\,\ref{Fig:BDparMK}b.  An orthogonal regression gives

\begin{equation}
\IeffK/\LKSun = 10^{-4.78 \pm 0.11} (\LKBul/10^{10}\LKSun)^{0.32 \pm 0.19}.
\label{Eqn:mueffserMK}
\end{equation}

\noindent ($S_\mathrm{R} = 0.34$; \Pnull\ $=0.16$).  Here and in following subsections, $K$-band surface brightness are derived from $H$-band values using $H-K=0.23$.  The distribution is nearly flat, as are the trends of $\mu_e$ with the disk luminosity \LKDis\ (Fig.\,\ref{Fig:BDparDisMK}b; $S_\mathrm{R} = 0.41$; \Pnull\ $=0.087$), with $\log(\sigma_{0})$ (Fig.\,\ref{Fig:BDparSIG}b; $S_\mathrm{R} = -0.48$; \Pnull\ $=0.065$), and with T (Fig.\,\ref{Fig:BDparT}b; $S_\mathrm{R} = 0.43$; \Pnull\ $=0.069$). 

\subsubsection{Central surface brightness}
\label{Sec:BulgeMu0}

Of key interest are the trends of central surface brightness with luminosity.  If bulge surface brightness profiles were homologous, i.e., if they had the same profile shapes, the trends found in \S\ref{Sec:MuEff} for \meanmue\ and $\mueff$ would imply central densities nearly independent of luminosity.  However, bulge profiles are not homologous, and, hence, peak surface brightness needs to be measured closer to the centers (Binggeli \& Cameron 1991).  

Even with the high spatial resolution of the \HST, the various approaches that may be envisaged to define and measure central surface brightness lead to different results.  Each approach has its own merits.  One measure is provided by the inwardly-extrapolated central value of the bulge S\'ersic profile, \muZeroSer.  Using an extrapolated central value is common practice for exponential and other profile models that have finite central density, and so we use this approach for our S\'ersic spheroids.  As compared to a direct reading of the $\mu(r)$ profile, \muZeroSer\ avoids the biases introduced by additional components present in the galaxy nuclei (Paper~III), and hence may be seen as a more accurate estimate of the central surface brightness of the bulge component.  We will show here that \muZeroSer\ shows a strong, monotonic trend with luminosity for bulges, as previously known for dwarf and giant ellipticals.   

Figure\,\ref{Fig:BDparMK}c shows $\muZeroSer \equiv -2.5\,\log\,\IZeroSer$ against the bulge absolute magnitude.  The distribution shows that the central surface brightness of bulges is a steep function of luminosity (compare to Fig.\,\ref{Fig:BDparMK}b).  After offsetting \muZeroSer\ ($H$-band) to the $K$-band with $\muZeroSerK=\muZeroSer - 0.23$, 
an orthogonal regression to the data points gives

\begin{equation}
\IZeroSerK/\LKSun = 10^{-3.37 \pm 0.15} (\LKBul/10^{10}\LKSun)^{0.81 \pm 0.33}.
\label{Eqn:mu0serMK}
\end{equation}

\noindent ($S_\mathrm{R} = 0.49$; \Pnull\ $= 0.044$).  
The relation is close to the maximum slope that is compatible with an increase of \reff\ with spheroid luminosity (given $L_{Sersic} = I(0)\,\reff^2\,2\pi\,n\,\Gamma(2n)/b_n^{2n}$).  
There is no sign of a turn-over at high luminosities.  
Indeed, if $n$ increases with luminosity, then the {\sl effective} surface brightness turn-over at bright magnitudes does not imply a turn-over of {\sl central} surface brightness (Binggeli \& Jerjen 1998; Graham \& Guzm\'an 2003).  Rather, $\mu_0$ continues to rise with luminosity due to the higher $n$ from higher luminosity spheroids.   
The dependence of $\muZeroSer$ on T is not significant (Figure\,\ref{Fig:BDparT}c; \Pnull\ $= 0.97$), indicating that Hubble type index does not determine central galaxy density.  Similarly, \muZeroSer\ does not correlate with the disk luminosity (Figure\,\ref{Fig:BDparDisMK}c; \Pnull\ $= 0.57$).

In Figure\,\ref{Fig:mu0MB}a we plot the S\'ersic-extrapolated central surface brightness \muZeroSer\ of our bulges against absolute magnitude, together with ellipticals from Caon et al.\ (1993) and dwarf ellipticals from Binggeli \& Jerjen (1998) and Graham \& Guzm\'an (2003). The values of \muZeroSer\ from bulges follows a general trend of brighter \muZeroSer\ for more luminous objects, common to spheroids of vastly different luminosities: objects plotted in Figure\,\ref{Fig:mu0MB}a include 
dwarf ellipticals and giant, 'core' ellipticals present in the Caon et al.\ sample.  (At the faint end, bulges are brighter than dEs of the same luminosity;  we argued in Paper~III that those nuclear disk components that do are not identified by our profile decomposition, might contribute to this offset.)

The quantity \muZeroSer\ is a useful parametrization of a {\sl bulge's} central surface brightness, but it can differ from the {\sl galaxy's} central surface brightness, to which the nuclear components also contribute.  The latter quantity is relevant as a diagnostic of dissipation during galaxy formation, and, e.g., for inferences of the nuclear evolution during galaxy mergers.  Because most of the nuclear components are unresolved even by \HST\ (Paper~III), the true galaxy's central surface brightness is generally not accessible from the present data.  We have therefore explored the trends defined by various measurements of $\mu(r)$ at a fixed, small angular radius; linear radius; or a fraction of the bulge's \reff.  The various measurements yield the same trends, namely, a brightening with increasing bulge luminosity. As an example, Figure\,\ref{Fig:mu0MB}b shows the surface brightness measured at a radius of 50 pc against absolute magnitude for our bulges, together with those of ellipticals (Lauer et al.\ 1995, hereafter L95; Caon et al.\ 1993), and dwarf ellipticals (Binggeli \& Jerjen 1998, Virgo cluster data; Graham \& Guzm\'an 2003, Coma cluster)\footnote{Surface brightness values from the literature have been scaled to the $H$-band as follows: for ellipticals, we use $V-H$ values derived from NED (http://nedwww.ipac.caltech.edu/) when available, otherwise we set $V-H=3.0$.  For dwarf ellipticals, we use $V-H=2.5$, which corresponds to a stellar population of metallicity 0.4 times Solar, age 5 to 10 Gyr and Salpeter IMF, from Vazdekis et al.\ (1996).}.  The choice of radius is driven by the angular resolution of WFPC2 and the distance to the Coma cluster.  
The distribution in Figure\,\ref{Fig:mu0MB}b is similar to those of e.g.\ Phillips et al.\ (1996, their Figure\,6);  Faber et al.\ (1997, hereafter F97, their Figure\,4); Graham \& Guzm\'an (2003, their Figure\,9).  Both for bulges and 'power-law' ellipticals, brightness increases monotonically with luminosity, the trend only breaking for the 'core' ellipticals due to the flattening of their profiles in the inner 100-200 pc (L95; Graham et al.\ 2003; Trujillo et al.\ 2004). We find a similar trend of brighter surface brightness with luminosity for any measurement of $\mu_{\rm cen}$ down to the resolution limit of our data, 10--20 pc.  

\subsubsection{Bulge concentration}
\label{Sec:BulgeN}

Figure\,\ref{Fig:NvsTvsBDvsMK} displays the behavior of the bulge S\'ersic index $n$, a measure of how centrally concentrated the stellar distribution is (Trujillo et al.\ 2001b) --- as a function of galaxy Type, $\log(B/D)$, and \MagKBul. We find values of $n$ from 0.7 to 3.1, a  lower range than typically reported from fits to bulge surface brightness profiles of early-type disk galaxies using ground-based data (APB95; Khoshroshahi et al.\ 2000; M\"ollenhoff \& Heidt 2001).  Paper~II discusses that the higher values of $n$ derived from ground-based data come from the inclusion of light from additional, distinct, central components, which forces the S\'ersic index up.  
Once the central components are dealt with, the {\sl bulge} light distribution shows values of $n$ lower than the classical de Vaucouleurs \r14\ behavior. 
The value of $n$ increases toward earlier types and toward higher values of $B/D$ (Fig.\,\ref{Fig:NvsTvsBDvsMK}a,b), as found by previous works (APB95; G01; Trujillo et al.\ 2001a).  

Virgo cluster ellipticals from Caon et al.\ (1993), and Coma cluster dwarf ellipticals from the \HST-based study of Graham \& Guzm\'an (2003) are included in Figure\,\ref{Fig:NvsTvsBDvsMK}c\footnote{A constant color term $F606W - K = 2.7$ was used to scale the Graham \& Guzm\'an dE absolute magnitudes, which corresponds to a stellar population of metallicity 0.4 times Solar, age 5 to 10 Gyr and Salpeter IMF, from Vazdekis et al.\ (1996).}, as well as late-type bulges from de Jong (1996) and analyzed by G01.  Our bulges follow the sequence defined by the Caon et al.\ ellipticals, while the later-type bulges of G01 are offset to lower values of $n$.  The reason for this offset is unclear; it could be related to the differences in viewing angles between the face-on de Jong sample and our inclined sample (see also M\"ollenhoff et al.\ 2006).  Also, it is plausible that our S\'ersic fits of the four lowest luminosity, inclined bulges are affected by undetected inner exponential components --- raising the bulge shape index above its true value.    

The dwarf ellipticals also have overall higher values of $n$ than the late-type bulges of similar magnitude.  This difference appears also for the dE sample of Binggeli \& Jerjen (1998) and is probably real.  The offset is not due to an incorrect color correction for the dEs; a lower metallicity would move the dE total luminosities faint-ward, while solar metallicities would make them brighter by only $\sim$0.3 mag.  It is unlikely that nuclear sources in the dEs are biasing the S\'ersic fits, as Graham \& Guzm\'an allowed for additional nuclear components in their fits, as we have done with the bulges sample. 

\subsubsection{The Faber-Jackson relation for Bulges}
\label{Sec:BulgeFJ}

We show the Faber \& Jackson (1976) $L-\sigma_0$ relation (see also Poveda 1961; Minkowski 1962; Fish 1964) for bulges in Figure\,\ref{Fig:BDparMK}f.  The velocity dispersions are aperture-corrected values from Ca{\sc II} triplet spectra (Falc\'on-Barroso et al.\ 2002).   We find a tight correlation which yields

\begin{equation}
\sigma_0 = 10^{2.04 \pm 0.04}( \LKBul/10^{10}\LKSun)^{0.35 \pm 0.05},
\label{Eqn:FJ}
\end{equation}

\noindent ($S_\mathrm{R} = -0.84$; \Pnull\ $= 5.6\cdot 10^{-4}$) giving a relation of $L \sim \sigma_0^{2.9 \pm 0.5}$.  Our exponent is lower than that found for luminous ellipticals: Faber \& Jackson give $L\sim \sigma^4$, while Pahre et al.\ (1998) give $L_K \sim \sigma_0^{4.1 \pm 0.2}$ for Coma ellipticals. The latter relation is plotted in Figure\,\ref{Fig:BDparMK}f, (dashed line).  
Formally the two relations differ.  However, the slope for Pahre et al.'s sample is largely determined by objects in the luminous range $M_K < -22$ mag, where few bulges lie.  
The deviations onset at $M_K > -22$ mag, where the Pahre et al.'s data show much scatter (see Fig.\,2 of Falc\'on-Barroso et al.\ 2002).  In our domain, the slope defined by bulges is very well constrained despite our small sample size.  It is plausible that, if Pahre et al.'s dispersions come from major axis spectra,  rotation might lead to overestimating the velocity dispersions.  We note that the FP of our bulges sample is very close to the J\"orgensen et al.\ (1996) FP of Coma ellipticals (Falc\'on-Barroso et al. 2002).  

The shallower slope we find is in agreement with the studies of faint elliptical galaxies with comparable luminosities.  Tonry (1981) repored a faint-end slope of $\sim$3, while Davies et al.\ (1983) and Held et al.\ (1992) report a slope of $\sim$2.5 for ellipticals fainter than -20 $B$-magnitudes and dwarf ellipticals, respectively.  The recent analysis of Coma faint early-type galaxies by Matkovi\'c \& Guzm\'an (2005) gives a value of $2.01 \pm 0.36$.  

Interestingly, central velocity dispersion shows no dependency with galaxy T (Fig.\,\ref{Fig:BDparT}f; $S_\mathrm{R} = -0.38$; \Pnull\ $= 0.11$), indicating that the bulge central potential does not vary along the early-to-intermediate Hubble sequence, except, perhaps, for our latest  (Sbc) galaxies.  Hence, along this part of the Hubble sequence, Hubble type is not determined by the depth of the central potential of the galaxy.

\subsection{Disk parameters}
\label{Sec:DiskParameters}

Our analysis of disk parameters differs from similar studies in the literature, in that we additionally focus on the scaling of disk parameters with {\sl bulge} luminosity (Figs.\,\ref{Fig:BDparMK}d,e).  Such relations may provide useful tests of hierarchical galaxy formation models (e.g.\ Cole et al.\ 2000; Navarro \& Steinmetz 2000), given that, in those models, disks largely grow around pre-existing bulges of merger origin.  To facilitate comparison with other works, we also show the trends of disk parameters with disk absolute magnitude (Figs.\,\ref{Fig:BDparDisMK}d,e), with central velocity dispersion (Figs.\,\ref{Fig:BDparSIG}d,e), and on Hubble type T (Figs.\,\ref{Fig:BDparT}d,e). Most disk parameters scale with bulge luminosity, and do not with Hubble type, as noted previously by others.  Perhaps more interestingly, we will show that, for the types analyzed in the present study, \textsl{disk parameters show weak or null correlations with $K$-band disk luminosity}. 

The disk scale length $h$ increases with bulge luminosity (Fig.\,\ref{Fig:BDparMK}d) as 

\begin{equation}
h/{\rm kpc} = 10^{0.35 \pm 0.04}( \LKBul/10^{10}\LKSun)^{0.36 \pm 0.08}.
\label{Eqn:hMK}
\end{equation}

\noindent ($S_\mathrm{R} = 0.84$; \Pnull\ $= 5.1\cdot 10^{-4}$).  It has a similar dependency on bulge luminosity than the bulge $\reff$ does, which leads to the result that the ratio $\reff/h$ must be independent of  bulge luminosity, which we discuss in \S\ref{Sec:BDscaling}.  

As expected, $h$ increases with disk luminosity \LumKDis\ (Fig.~\ref{Fig:BDparDisMK}d).  However, the relation, 

\begin{equation}
h/{\rm kpc} = 10^{0.034 \pm 0.111}( \LKDis/10^{11}\LKSun)^{0.55 \pm 0.14}.
\label{Eqn:hMKDis}
\end{equation}

\noindent is weaker ($S_\mathrm{R} = 0.62$; \Pnull\ $= 1.1\cdot 10^{-2}$) than found against \LumKBul.  

The dependence of $h$ with Hubble type index T is shown in  Figure\,\ref{Fig:BDparT}d.  Disk scale lengths show no trends with T ($S_\mathrm{R} = 0.16$; \Pnull\ $= 0.49$), a fact already noted by Freeman (1970).  This indicates that the increase of $h$ with bulge luminosity (Fig.\,\ref{Fig:BDparMK}d; eqn.\,\ref{Eqn:hMK}) is not a consequence of earlier-type galaxies in our sample being intrinsically larger galaxies.  Disk scale lengths are also rather constant with Hubble type in the samples of HPG03 ($H$-band data) and of de Jong (1996, $K$-band), which are overall later type than ours.  

The disk face-on central surface brightness $\muZeroDis \equiv -2.5\,\log\,\IZeroDis$ (Fig.\,\ref{Fig:BDparMK}e) also scales with bulge luminosity:  

\begin{equation}
\IZeroDisK/\LKSun = 10^{-5.55 \pm 0.06} (\LKBul/10^{10}\LKSun)^{-0.28 \pm 0.09},
\end{equation}

\noindent i.e., a gentle but well-defined faintward trend with increasing bulge luminosity ($S_\mathrm{R} = -0.60$; \Pnull\ $= 0.013$).  For reference, the Freeman (1970) canonical disk central surface brightness is shown (horizontal dashed line).  The disk $\mu_0$ does not vary with T over our range of Hubble types (Fig.\,\ref{Fig:BDparT}e; $S_\mathrm{R} = 0.13$; \Pnull\ $= 0.57$).   G01 and HPG03 find a similar behavior, together with a mild decrease for later types (T$>$4).  

The disk central surface brightness is shown against disk absolute magnitude in Figure\,\ref{Fig:BDparDisMK}e.  We see no correlation, and one may be tempted to see this distribution as a manifestation of Freeman's law; in fact, comparison with Figure\,\ref{Fig:BDparMK} (\muZeroDis\ against \MagKBul) shows that a second parameter, namely \MagKBul, is responsible for the width of the distribution, and that the lack of a correlation is solely a consequence of a choice of independent variable, \MagKDis, that scrambles the distribution of \muZeroDis\ against \MagKBul.  

\subsection{Scaling of bulge and disk parameters}
\label{Sec:BDscaling}

Figures\,\ref{Fig:BDparMK}g-i,\,\ref{Fig:BDparSIG}g-i, and \ref{Fig:BDparT}g-i show how the bulge-to-disk ratios of central brightness, spatial scales and luminosities depend on the bulge absolute magnitude, central velocity dispersion, and Hubble type index T, respectively.  The dramatic rise of 
$\log(\IZeroBul/\IZeroDis)$ with bulge luminosity ($10^1 - 10^3$, Fig.\,\ref{Fig:BDparMK}g; $S_\mathrm{R} = -0.65$; \Pnull\ $= 7.6\cdot 10^{-3}$) is dominated by the brightening of the bulge central intensity with increasing bulge S\'ersic index $n$.  In contrast, the position along the Hubble sequence has a weak relation with 
$\log(\IZeroBul/\IZeroDis)$: figure\,\ref{Fig:BDparT}g shows a big scatter with a drop at only the latest (Sbc) type ($S_\mathrm{R} = -0.13$; \Pnull\ $= 0.57$).  A null trend is found also with disk absolute magnitude (Fig.\,\ref{Fig:BDparDisMK}g; \Pnull\ $=0.42$).

The ratio of spatial scales $\reff/h$ is shown against bulge absolute magnitude, central velocity dispersion and T in Figures\,\ref{Fig:BDparMK}h, \ref{Fig:BDparSIG}h, \ref{Fig:BDparT}h, respectively.  $\reff/h$ shows no definite trend with any of these parameters: \Pnull\ = 0.41, 0.76, and 0.48 for \MagKBul, 
$\log(\sigma_{0})$ and T, respectively.  
Similar dependencies of $\reff/h$ with $n$ and with T have been found for later-type spirals (de Jong 1996; G01; HPG03).  The non-dependence of $\reff/h$ with T is the basis for the statement by Courteau et al.\ (1996) that the Hubble sequence is scale-free, a topic we address in \S\ref{Sec:DiscussionHubbleSequence}.  

Finally,  as expected, $B/D$ increases with bulge-luminosity (Fig.\,\ref{Fig:BDparMK}i).  The relation is 

\begin{equation}
B/D = 10^{-0.59 \pm 0.10}( \LKBul/10^{10}\LKSun)^{0.62 \pm 0.15},
\label{Eqn:BDMK}
\end{equation}

\noindent ($S_\mathrm{R} = -0.70$; \Pnull\ $= 3.7\cdot 10^{-3}$).  The increase in $B/D$ is dominated by the increase in bulge luminosity, as already pointed out by Trujillo et al.\ (2002).  $B/D$ increases by a factor of nearly 100 over 8 mag in \MagKBul. In contrast, a null trend is found with disk absolute magnitude (Fig.\,\ref{Fig:BDparDisMK}i; \Pnull\ $=0.86$).

Equation \ref{Eqn:BDMK} implies that disk luminosities increase as $L_\mathrm{Disk} \sim {L_\mathrm{Bulge}}^{0.38}$.  
$B/D$ also increases with $\log(\sigma_{0})$ (Fig.\,\ref{Fig:BDparSIG}i; 
$S_\mathrm{R} = 0.69$; \Pnull\ $= 6.2 \cdot 10^{-3}$), 
and $B/D$ depends only weakly on T (Fig.\,\ref{Fig:BDparT}i; 
$S_\mathrm{R} = -0.48$; \Pnull\ $= 0.041$, 
driven by the drop at T = 4), and with considerable scatter (see also de Jong 1996), highlighting the problem of using T-based $B/D$ ratios (e.g., Baggett, Baggett, \& Anderson 1998).  Although $B/D$ appears to strongly decrease for T$>$3, it is rather constant for Sb and earlier galaxies.  

\section{Discussion}
\label{Sec:Discussion}

\subsection{The trend of central surface brightness with bulge luminosity}
\label{Sec:DiscussionMu0}

Whether the central surface brightness of spheroidal components becomes brighter or fainter with increasing spheroid luminosity provides useful diagnostics on galaxy formation.  The sign of this correlation results from the combined effects of dissipation (e.g., Kormendy 1989), feedback from star formation (e.g., Silk 2005), and dynamical heating by merging black holes (e.g., F97).  
In mergers between unequal galaxies, it determines which component ends up populating the center of the merger remnant (Zurek et al.\ 1988; Balcells \& Quinn 1990).  
Our trends of $\mu_0$ with luminosity (Fig.\,\ref{Fig:mu0MB}) show that both the inwardly-extrapolated bulge $\mu_0$, and $\mu$ measured at a fixed radius, become brighter with luminosity (\S\ref{Sec:BulgeMu0}, Fig.\,\ref{Fig:BDparMK}c, Fig.\,\ref{Fig:mu0MB}).  We have verified that $\mu$ measured at $r=20$ pc from the center, and at a fixed fraction of $\reff$, specifically 0.2\,$\reff$, follow the same behavior.  Furthermore, the slope defined in Figure\,\ref{Fig:mu0MB}b by our bulges (brighter $\mu_0$ for more luminous objects) becomes steeper the closer to the center we measure $\mu$.  Jergen et al.\ (2000) have shown that, at the faint end of spheroid luminosity, the highly resolved, and also common, dwarf galaxies around the Milky Way and M31 have central profiles that do not show central excesses over their outer S\'ersic profile fits and have very low central surface brightness.  
The trend of brighter central surface brightness for higher luminosities is therefore well established for spheroids of vastly different luminosities, from giant ellitpicals to dEs (Fig.\,\ref{Fig:mu0MB}a). 

Our findings contrast with the arguments of F97 who favor of a trend of {\sl fainter} central surface brightness with increasing luminosity for ellipticals and bulges.  F97 reasoned that, if M32 resided in the Virgo Cluster, its exceedingly high central surface brightness would be smoothed by the PSF to values close to those of other low-luminosity Virgo E's.   By the same argument, bulges and low-luminosity ellipticals at distances comparable to Virgo might have M32-like central profiles which, with sufficient spatial resolution, would yield $\mu_0$ values following  a faint-ward extrapolation of the trend defined by giant ellipticals, with $M_B + 5\,\log h_{50} \la -20.5$ mag, in Figure~\ref{Fig:mu0MB}b.  

Although Jerjen et al.\ (2000) have shown that the picture described by 
F97 breaks down at the low-luminosity end, one could argue that the central {\sl galaxy} surface density, 
rather than {\sl bulge} surface density, of some nucleated, intermediate-luminosity  
galaxies and bulges may follow the trend argued for by F97 (unresolved components [PS] reside in the centers of 58\% of our bulges, see Paper~III).
Specifically, if the central surface brightness of the PSs, $I_{0,\mathrm{PS}}$, became sufficiently {\sl brighter} for fainter PS (and hence bulge) luminosities, the trend between bulge luminosity and {\sl galaxy} central surface brightness could in principle have an opposite slope to that shown in Figure\,\ref{Fig:mu0MB} for the trends between bulge luminosity and {\sl bulge} central surface brightness.  

We find this unlikely. Both the central surface brightness of the underlying S\'ersic component ($\IZeroSer \sim (\LKBul)^{0.81}$, eqn.\,\ref{Eqn:mu0serMK}), 
and the PS luminosities 
($\LumPt \sim (\LKBul)^{0.91}$, Paper~III)
rapidly become fainter toward lower luminosities, requiring exceedingly high densities for the unresolved components (over 5 magnitudes in surface brightness) in order to reverse the trend of Figure\,\ref{Fig:mu0MB}.  B\"oker et al.\ (2004), using \HST/PC imaging,  find that the sizes of nuclei do not correlate with nuclei luminosities, hence the central surface brightness \textsl{of the nuclear clusters} $I_{0,\mathrm{PS}}$ does not correlate with \LumPt\  (for a sample of late-type spiral nuclei.)   

The compact elliptical galaxy M32 does provide a counter-example to the previous reasoning.  The ground-based $R$-band surface brightness profile of Kent (1987; see Graham 2002), and the color profiles from Peletier (1993), yield $\mu_H$(50\,pc)=14.64 mag.  Given the absolute magnitude of $M_B=-15.74$ for the M32 bulge\footnote{We assume a bulge-to-disk ratio of $\log(B/D)=0.22$ from Graham (2002); we use the apparent blue magnitude from LEDA, and a distance modulus of 24.43 (Jacoby et al.\ 1992).  Ignoring the bulge-disk decomposition in determining the absolute magnitude would not affect our argument.}, M32 lies within the sequence of bulges and low-luminosity ellipticals in Figure\,\ref{Fig:mu0MB}.  Yet, the \HST\ profiles in F97 yield $\mu_H$(0.1\arcsec)=8.52 mag, i.e., five magnitudes above bulges and ellipticals of the same luminosity.  Note that the M32 S\'ersic profile shape index derived by Graham (2002) is a modest $n=1.5$, well within the range for bulges of similar luminosities; the high central brightness that makes M32 a prototype, compact elliptical galaxy really refers to a region of radius smaller than 37 parsec only (Graham 2002 excluded radii smaller than 10\arcsec\ from his fit). This inner region is M32's  'unresolved source', using the term applied to our bulges.  If the unresolved sources in bulges had similar structure to the M32 nucleus, they could reach very high central surface brightnesses despite their low luminosities.   
However, it is unclear whether M32 is representative of spheroids such as those analyzed in this paper; the bulge of M31 might be a more adequate model; it
does not show such high central surface brightness.  

\subsection{Bulge luminosity as a measure of the structure of disk galaxies}
\label{Sec:DiscussionHubbleSequence}

Of the four sets of correlations involving bulge luminosity, disk luminosity, bulge central velocity dispersion, and galaxy Hubble type T, relations involving bulge luminosity show the strongest level of significance.  This was perhaps expected for bulge intrinsic parameters, but it is significant that disk parameters and bulge-to-disk ratios correlate with bulge luminosity as well.  They also do so with the central velocity dispersion $\sigma_{0}$, as a result of the correlation between $\sigma_{0}$ and luminosity for bulges.   In contrast, the dependencies with T are weak or non-existent.  

Such lack of scaling with T imply a stronger form of the statement from de Jong (1996) and Courteau et al.\ (1996) that the Hubble sequence is scale-free.  These authors noted that bulge and disk scale lengths are correlated, and that their ratio $\reff/h$ is independent of Hubble type.  We reproduce those results in our sample, as do HPG03 in theirs.  Here, however, the lack of dependence on Hubble type is not restricted to length scales: the distributions of {\sl most} bulge, disk, and bulge-to-disk parameters versus T shows an absence of trends (Fig.\,\ref{Fig:BDparT}) for our early-type disk galaxy sample.  HPG03 find a similar absence of correlations for a later-type sample, selected with different criteria (galaxies belonging to the Perseus-Pisces supercluster), and analyzed with a different (2D) code, suggesting that this behavior may be common for bulge-disk galaxies.  

Although the Hubble sequence is scale-free, galaxies themselves are not. 
The bulge luminosity (and velocity dispersion) correlate well with the
other properties of the bulge, the disk, and the scaling of these two 
components.

What structural parameters define the Hubble sequence?  Focussing on the early-to-intermediate disk galaxies covered by our sample (S0--Sbc), structural changes with T are only found in the latest, Sbc galaxies (lower B/D,  lower $I_{0,Bul}/I_{0,Disk}$ and marginally lower central velocity dispersions).  For types S0 to Sb, Hubble type must be  dominantly given by the spiral pattern (Block \& Puerari 1999; Seigar, Chomey, \& James 2003), dust content, and star formation activity.  Furthermore, the latter must not be affected by the spheroid luminosity or mass, nor by the relative sizes and brightness of disk and bulge. It appears that \textit{the luminosity and size trends commonly associated with the definition of the Hubble sequence appear when extreme late-type galaxies are compared to early-types; but those trends are not gradual, and are absent among early-to-intermediate disk galaxies}.  

That bulge luminosity strongly correlates with the properties of both the disk and the bulge-to-disk ratios may be expected from galaxy formation scenarios in which the bulge precedes the disk.  This includes monolithic collapse models akin to the Eggen, Lynden-Bell, \& Sandage (1962) model for the formation of the Milky Way, or CDM-based hierarchical galaxy formation scenarios in which disks grow around already formed bulges of merger origin (e.g.\ Baugh et al.\ 1996; Abadi et al.\ 2003; Sommer-Larsen,  G{\" o}tz, \& Portinari 2003).  Our results are probably also compatible with the inside-out galaxy formation model of van den Bosch (1998).  The early formation of bulges does not necessarily imply a classical, slowly-rotating spheroid structure for bulges; if star formation timescales exceed the dynamical time, such bulges might show rapid rotation and some of the disky properties highlighted by Kormendy (1993).  Are our results compatible with secular evolution models in which the bulge grows from instabilities in an already formed disk?  The scaling of bulge and disk parameters is generally taken to support secular evolution models (Courteau et al.\ 1996; Zhang 2004).  
Here, we find that most bulge parameters show non-significant correlations with the disk $K$-band luminosity, suggesting that bulge properties (\reff, mass, central density) do not know how massive is the host disk.  Perhaps more puzzling, the disk parameters show more well-defined trends with bulge luminosity than with disk luminosity.  Indeed, only one galaxy parameter, namely the disk scale length, correlates with the disk luminosity ($\LumKDis \sim h^{\sim 2}$ (eqn.~\ref{Eqn:hMKDis}).  Explaining these relations, or lack of thereof, between galaxy structural parameters, will remain key challenges for secular evolution models.  

This difficulty is compounded by the evidence of stellar population ages.  We show in Paper~I that the bulge populations in the present sample are as old as cluster ellipticals.  Our measurements were performed away from the disk plane, hence ongoing or episodic star formation of secular origin could take place in inner regions of the disk. But  old population ages are inferred for the nuclei of later-type spirals from NIR spectroscopy (Bendo \& Joseph 2004).  To be compatible with this evidence, 
secular evolution must have been of little relevance for the growth of the bulge, or else confined to look-back times of order 10 Gyr for our galaxies: the situation essentially reduces to the van den Bosch model; and, timescales need to be known to verify that the term 'secular' applies.  The situation may be different for bulges of later types, and of barred galaxies, which are not addressed in this paper (see, e.g., Fathi \& Peletier 2003; Castro-Rodr\'\i guez \& Garz{\' o}n 2004);  
although, in late-type bulges, young populations are statistically associated with morphological disturbances, pointing to external, rather than internal-secular, processes (Kannappan et al.\ 2004).  

Hence, from the population information in Paper~I and the structural information in this paper, it appears that  bulges of unbarred, early- to intermediate-type disk galaxies cannot have significantly grown from disk instabilities in the last $\sim$10 Gyr.  
A small amount of star formation is probably occurring in almost every galaxy nucleus, as evidenced by dust in the very centers of bulges (e.g.\ Paper~I), and by the presence of young stars (Freeman \& Bland-Hawthorn 2002).  Bar dynamics could be feeding such activity, but, for early-type disk galaxies, star formation level is too low to affect the general bulge population.  Our conclusions on bulge formation/growth are consistent with the current understanding that many 'mature' galaxies were largely in place by $z=1$ (e.g.\ Brinchmann \& Ellis 2000; Simard et al.\ 2002), and that early-type galaxies in the field became red between redshifts $z=1$ and 2 (Eliche-Moral et al.~2006).  

\acknowledgements{This research has made use of the NASA/IPAC Extragalactic Database (NED) which is operated by the Jet Propulsion Laboratory, California Institute of Technology, under contract with the National Aeronautics and Space Administration.  This research has made use of the HyperLeda database.  The United Kingdom Infrared Telescope is operated by the Joint Astronomy Centre on behalf of the U.K. Particle Physics and Astronomy Research Council. }

\end{document}